# Dimming of the 17th Century Sun

Peter Foukal[1], Ada Ortiz[2], and Roald Schnerr[3]

[1] Heliophysics, Inc., Nahant, Massachusetts, USA 01908,

[2] Institute of Theoretical Astrophysics, University of Oslo, PO Box 1029 Blindern, N 0315 Oslo, Norway,

[3] Institute for Solar Physics, Royal Swedish Academy of Sciences, Alba Nova University Center, Stockholm;

Dept. of Astronomy, Stockholm University, Alba Nova University Center, 10691 Stockholm, Sweden.

**Abstract**

Reconstructions of total solar irradiance (TSI) rely mainly on linear relations between TSI variation and indices of facular area. When these are extrapolated to the prolonged 15th - 17th century Spörer and Maunder solar activity minima, the estimated solar dimming is insufficient to explain the mid- millennial climate cooling of the Little Ice Age. We draw attention here to evidence that the relation departs from linearity at the lowest activity levels. Imaging photometry and radiometry indicate *an increased TSI contribution per unit area from small network faculae* by a factor of 2-4 compared to larger faculae in and around active regions. Even partial removal of this more TSI – effective network at prolonged minima could enable climatically significant solar dimming, yet be consistent with the weakened but persistent 11- yr cycle observed in Be 10 during the Maunder Minimum. The mechanism we suggest would not alter previous findings that increased solar radiative forcing is insufficient to account for 20th century global warming.

## 1. Introduction

Studies over the past 30 years have shown that sunspots decrease the total solar irradiance (TSI), while the more fragmented magnetic structures called faculae increase it (see e.g. Fröhlich and Lean 2004; Solanki and Krivova 2004; Foukal et al. 2006 for recent reviews). The dimming by spots can be modeled with sufficient accuracy for studies of solar impacts on climate, to the beginning of the spot record in the early 17th century.

The TSI effect of the faculae is harder to estimate. The approximately 200 km diameter of the facular magnetic flux tubes is barely resolvable (e.g. Keller 1992; Berger et al. 1995) so area measurements are uncertain. Also, even the sign of their contrast depends on diameter (e.g. Lawrence et al. 1993; Ortiz et al. 2006; Spruit and Zwaan 1981), on wavelength (e.g. Lawrence et al. 1993: Foukal et al.2004), and on viewing angle (e.g. Muller 1975; Ortiz et al. 2006)).

Consequently, TSI reconstructions extending long enough for climate modeling have relied mainly on regressions between the facular TSI contribution and proxies of their area such as the F10.7 microwave flux or the Mg II index (e.g. Foukal and Lean 1988; Solanki and Fligge1998; Fröhlich 2009)). Linear relations derived using space – borne radiometry of TSI available since 1978 have been used to reconstruct the TSI variation caused by spots and faculae back to the beginning of the spot record.

When these are used to reconstruct TSI during the prolonged 17$^{th}$ century Maunder Minimum of activity, the decadally - smoothed solar dimming caused by the disappearance of spots and active region faculae is barely 0.04 %. Such a small change in solar radiative forcing is insufficient by a factor 3-5 to explain the mid- millennial climate cooling known as the Little Ice Age (Wigley and Raper 1990; Foukal et al. 2006). Evidence for a larger dimming caused by disappearance of *all* photospheric magnetism, was put forward by Lean et al. (1992). Using a linear relation between facular area and TSI change, they estimated an additional dimming of roughly 0.2%, if even the smallest network and intra-network elements of photospheric magnetism were to disappear.

The proposed gradual solar brightening through the 20$^{th}$ century was, however, in conflict with archival solar images (Foukal and Milano 2001). Also, the evidence for disappearance of all photospheric magnetism was based on an interpretation of stellar spectrophotometry that later required revision (see Foukal et al. 2004; Hall and Lockwood 2005 for a discussion).Nevertheless, it remains likely that network area *did* decrease during the Maunder Minimum, although radio- isotope measurements show a weakened 11- yr cycle (Beer et al. 1998), suggesting that some photospheric magnetism persisted.

In Sections 2 and 3 we review results from photometric imaging and radiometry which indicate that the relation between facular indices and TSI variation becomes *non – linear* at low activity levels. Specifically; the smaller – diameter facular elements found in the quiet magnetic network make a relatively greater contribution to TSI (per unit area) than the larger – diameter faculae encountered in and around active regions. In Sections 4 and 5 we consider how the removal of progressively smaller flux tubes during prolonged activity minima might dim the Sun more than expected from the linear models.

**2. The Size – Dependence of Facular Brightness Derived From Photometric Imaging**

Faculae are now recognized to be depressions in the photosphere whose bright walls shine directly into space, accounting for their enhanced brightness near the limb (Spruit 1976 ; Lites et al. 2004).Their disk – center excess continuum brightness ( ~ 0.1%) is barely detectable (Foukal and Fowler 1984; Lawrence et al. 1988), but these and other measurements at higher angular resolution (e.g. Muller 1975) suggested that faculae of all sizes were brighter than their surroundings, whatever their disk position.

Infrared detectors changed this picture. Worden (1975) showed evidence for dark facular structures at 1.6 microns, where we see deepest into the photosphere. Observations using the first IR arrays (Foukal et al.1990; Moran et al.1992), showed that faculae of magnetic flux > 2-3 x 10$^{18}$ Mx are dark near disk center, although they are bright in Ca K. Imaging at higher angular resolution showed that their darkness can be detected in visible continuum (e.g. Zirin and Wang 1992; Wang et al. 1998). Since then, several studies (e.g. Lawrence et al. 1993; Ortiz et al. 2006) have shown that only the smallest faculae, identified with the magnetic network, are bright across the whole disk. The chromospheric brightness of the dark faculae leads to their inclusion in direct and proxy (e.g. Ca K, F10.7, Mg II) measurements of facular area. Therefore, the increasing fraction of such dark flux tubes decreases the ratio of (TSI contribution) / (chromospheric facular area index) in faculae larger than the quiet network.

Skumanich et al.(1975) and Ortiz et al.(2006) reported a contrast increase with magnetic flux in the network. At their angular resolution of ~ 2 arc secs this reflects increasing filling factor of flux tubes in the instrument aperture. The facular flux tubes are all of ~ 1 kG magnetic intensity (Stenflo 1973). So the *magnetic* signal is proportional to the sum of their areas, and is insensitive to their size distribution.

Their *brightness contrast* is, however, dependent upon size distribution. Models (Spruit 1976) and realistic, extensively validated 3 – D simulations (e.g. Keller et al. 2004; Carlsson et al. 2004) predict that smaller flux tubes should be brighter per unit area. The main reason is that flux tube heating by radiation from the bright tube walls (which scales as the tube's diameter) becomes increasingly important relative to cooling by inhibition of convection along the tube's axis (which scales as its cross sectional area).

This is supported by the finding of Ortiz et al. (2006), that the contrast/B (i.e. *per unit of magnetic flux)* increases with decreasing flux. Their result refers to contrast measured using the MDI continuum data at 676.8 nm wavelength, but wide- band facular contrasts are likely to be similar since measured bolometric contrasts lie close to those around the peak of the H- opacity curve in the red (Foukal et al. 2004).

The Ortiz et al. result is based on *maximum* contrast measured on curves of contrast versus heliocentric angle, θ. But their data can be used to derive a similar result for the excess radiative flux, ΔF/F, of a facula using the standard photometric relation

$$\Delta F/F = 2\pi \int C(\mu) \, \mu \, d\mu \quad \quad \quad (1)$$

where $C(\mu) = (I_f(\mu) – I_p(\mu)) / I_p(\mu)$ is the contrast of a facula of intensity $I_f$, relative to photosphere of intensity $I_p$ at the heliocentric cosine $\mu = \cos \theta$.

Figure 1 shows this normalized excess radiative flux of individual faculae plotted against their magnetogram signal ranging from B < 90 G measured in quiet network faculae, to B = 500 G – 600 G in active region faculae. This signal is the mean magnetic intensity averaged over the magnetograph entrance aperture; the network magnetic signal < 90 G given above corresponds to a flux integrated over the aperture of 2 x $10^{18}$ Mx. We see in the upper panel that ΔF/F peaks at intermediate B values around 350 G. But in the lower panel we see that the ratio (1/B) (ΔF/F) *increases monotonically by about a factor of four between the strongest and weakest facular magnetic signals.*

This result indicates that the excess radiative flux (and TSI effect) of faculae is *not* simply proportional to their filling factor within the instrumental aperture. Removal of a facula will decrease TSI more if its magnetic signal B arises from a larger number of smaller flux tubes, than if it arises from a smaller number of larger ones. This has important implications for TSI reconstruction at times of low activity.

**3. Evidence From Radiometry**

Some additional support for an increased TSI – effectiveness of the network is suggested by recent analyses of TSI radiometry. The PMOD (Physical – Meteorological Observatory Davos) radiometric composite shows a TSI decrease during the 2007-2009 activity minimum to values which are 0.016 +/- 0.0086 % below those recorded in the two earlier minima (Fröhlich 2009).

This dip is greater than expected from the decrease in chromospheric proxies of facular TSI contribution (Foukal et al. 2009; Fröhlich 2009). Fig 2 shows this discrepancy using F10.7 as a *linear* proxy of the facular TSI contribution. Although the confidence level of this excess TSI dip is below 95%, Fröhlich has suggested (see also Steinhilber 2010; Ball et al.2011; Krivova et al. 2011 for discussion), that it might be caused by a general photospheric dimming outside of magnetic flux tubes. Our alternative explanation is that, with prolonged low solar activity around 2007 - 2009, the effect of the remaining flux tubes on TSI increased relative to their effect on chromospheric and transition radiations used in the proxies.

The increased TSI - effectiveness can be estimated by comparing the 20 % excess TSI dip to the (only) 5% excess decrease of a proxy index like F10.7 (Foukal et al. 2009; Fröhlich 2009). The reproducibility of the radiometry is uncertain at this level, but the discrepancy is consistent with a TSI contribution of the network about 2-4 times greater (per unit area) than the contribution of the enhanced network that determines the 11 – year TSI variation (Foukal and Lean 1988). This is about the same factor by which network flux exceeds active region facular flux in the photometry discussed in Section 2 above.

## 4. The Network Contribution to TSI

Increased brightness of network faculae is important to TSI reconstruction if they make a significant contribution to quiet Sun irradiance; this can be estimated as follows. The measured 11 – yr TSI modulation of ~ 0.06% amplitude (e.g. Frohlich and Lean 2004) arises largely from the solar cycle change in filling factor of the enhanced network (Foukal and Lean 1988). An estimate of the quiet network TSI contribution, ΔTSI, can then be obtained (assuming for now equal brightness of quiet and enhanced network) from:

ΔTSI ~ 0.06% x ($f/\Delta f_e$) …………………………………………………………………….. (2)

Here $f = A/2\pi R^2$ is the filling factor on the hemisphere of quiet network of area A, R is the solar radius, and $\Delta f_e$ is the change in filling factor of enhanced network between activity minimum and maximum. Both $f$ and $\Delta f_e$ are sensitive to angular resolution and magnetogram threshold, but their ratio is more stable.

Values obtained from MDI data (Ortiz et al. 2006), based on definition of quiet and enhanced network as pixels of B < 90G and 90 < B < 130 G respectively, yield ($f/\Delta f_e$) ~ 1.4 and ΔTSI ~ 0.09%, which agrees approximately with the (also linear) estimate by Lean et al. (1992). Our value may under- estimate ΔTSI, since $f$ obtained from the MDI data of relatively high magnetic threshold is lower than that obtained from either KPNO magnetograms (Foukal et al. 1991) or from Ca K images (Foukal and Milano 2001).

But even if this linear estimate were low, if the quiet network is 2 – 4 times brighter than the larger faculae in enhanced network and active regions, we would expect that ΔTSI should lie approximately in the range between (2-4) x 0.09%, therefore between 0.18% and 0.36%.

Spruit and Zwaan (1981) found that brightening of smaller flux tubes is accentuated in the wings of the Mg I b line which are formed in the high photosphere. So brightening would be even greater in the ultraviolet continuum formed mainly at these heights, than in the red continuum observed by Ortiz et al. (2006). This should further increase the TSI effect of the network.

In principle, the network TSI contribution is directly measurable from photometric and magnetograph imaging. The required observations would be photometry covering the broad wavelength range accepted by the radiometers, over the full range of viewing angle, with sufficient angular resolution (including magnetograms) to distinguish the smallest faculae. Observations meeting all these criteria do not yet exist.

Approximations are, however, available. Schnerr and Spruit (2011) find a value ~ 0.15% from high resolution observations near 630.2 nm with the Swedish Solar Telescope. Their measurements so far refer only to disk center; the value would be higher if measurements closer to the limb were used. It is instructive to compare their value with an estimate obtained from the lower – resolution MDI data of Ortiz et al. (2006).

The contribution, ΔTSI, of network faculae can be calculated from the relation (Foukal et al.1991)

$$\Delta TSI/TSI = f/2 \int C(\mu) (3\mu+2) \mu d\mu \ldots\ldots\ldots\ldots\ldots\ldots\ldots\ldots\ldots\ldots\ldots\ldots\ldots\ldots\ldots\ldots\ldots\ldots (3)$$

where the term $(3\mu + 2)$ expresses decreasing facular TSI contribution toward the limb due to photospheric limb darkening. Evaluation of this relation (3) using $f$ and $C(\mu)$ from MDI data at the 1996 solar minimum leads to a value ΔTSI/TSI ~ 0.01%. This is an order of magnitude lower than the value of Schnerr and Spruit.

These authors have shown that their estimate depends upon angular resolution; it is only about half as large when they use Hinode data of 0.32" angular resolution, as it is using the SST data of 0.16" resolution. One reason is that much of the oppositely – signed, small- scale *magnetic* structure cancels out at lower angular resolution. Another is that the *brightness* fluctuations are smoothed, so that more are averaged into the background and removed. These two effects probably explain why the MDI data of 2"x 2" angular resolution, and using a relatively high 50 G magnetic threshold, yield a much lower value.

This comparison illustrates why estimates of the network TSI contribution obtained using photometry and magnetograms that do not resolve the smallest solar flux tubes, must be considered as lower limits to the values of roughly 0.2% - 0.4% expected if the network is 2-4 times brighter than larger faculae.

**5. Implications for Solar Dimming at the Spörer and Maunder Minima.**

In the analysis of Lean et al.(1992) removal of all network during the Maunder Minimum would decrease TSI by ~ 0.1% below its value at normal sunspot minima. That is; $f$ would decrease from its solar minimum value $f$ ~ 0.15 (Foukal et al.1991; Foukal and Milano 2001) to $f$ = 0. If the network contrast is twice that of larger faculae, the same 0.1% decrease of TSI would require a decrease of $f$ by only 50% to $f$ = 0.08 instead of by 100%, to $f$ = 0.

The actual decrease of $f$ during the Maunder Minimum is unknown. But models of photospheric field evolution indicate a decay time for non – axisymmetric fields of 1-2 years (Wang et al. 2000). This does not take into account the possibility that network fields may be maintained by other mechanisms than just the decay of active regions (e.g. Schüssler & Vogler 2008). Although these mechanisms deserve closer study, a decrease in $f$ by 50% during the many decades – long Spörer and Maunder minima seems not unreasonable considering that the recent anomalous 5% decrease in facular area proxies occurred in response to the 2008 minimum that was prolonged by less than 2 years. Such a decrease in network area, but not a complete removal, might be consistent with radio-isotope evidence for a residual 11-year activity cycle during the Maunder Minimum (Beer et al. 1998).

As mentioned earlier, this 0.1% dimming would act in addition to the roughly 0.04% decrease of TSI caused by disappearance of the 11 year cycle in TSI. Climate modeling indicates that a TSI decrease of roughly 0.15 % during the very broad, $15^{th}$ – 19 th century solar activity decline might explain much of the terrestrial cooling described as the Little Ice Age (e.g. Wigley and Raper 1990; Foukal et al. 2006).

Solar activity minima between 1914 – 1996 exhibit no significant secular increase in $f$ (Foukal and Milano 2001). This argues against a secular increase of TSI due to increasing network area *during the $20^{th}$ century*, as proposed in addition to 11-year TSI modulation by Lean et al.(1995) and by Lockwood and Stamper (1999). This finding from archival solar images is supported by the subsequent reconsideration of such additional secular solar brightening over the past century (Lean, Wang and Sheeley 2002; see also Svalgaard and Cliver 2010).

**6. Conclusions**

a) Imaging in the near-infrared and visible spectral regions shows that Ca K faculae of magnetic flux greater than 2-3 x $10^{18}$ Mx tend to be *dark* in photospheric continuum except near the limb. Only the smallest faculae, observed in the quiet network and intra-network, are bright at all disk positions. Analysis of the excess facular radiative flux, $\Delta F$, and magnetic signal, B, in such measurements shows that *the ratio $(1/B)(\Delta F/F)$ increases with decreasing B.*

b) The decrease in B is generally ascribed to decreasing cross sectional area (rather than decreasing magnetic intensity) of the kilogauss flux tubes within the measuring aperture. So *brightness per unit area increases with decreasing flux tube area* progressing from the larger faculae in active regions to enhanced and quiet network. This agrees with flux tube models and simulations (e.g. Spruit 1976; Carlsson et al. 2004; Keller et al. 2004).

c) This brightening implies that *removal of network faculae decreases TSI proportionately more than removal of the same area of larger faculae* in and around active regions. Thus, the relation between the TSI contribution of faculae and their area is no longer linear at the lowest activity levels. The contrast per unit area of network appears to be greater than that of faculae in and around active regions, by a factor of approximately 2-4.

d) The TSI decrease reported by Fröhlich (2009) during the extended 2007 – 2009 activity minimum was approximately four times greater than can be explained using a linear relation between TSI and proxy indices of facular area. Fröhlich (2009) has suggested that, although its confidence level is marginal, the discrepancy might imply a general photospheric cooling. We suggest an alternative explanation that, *during the prolonged 2007 – 2009 minimum, the decrease in network area was about four times more effective in reducing TSI than expected in the linear models.*

e) An increased TSI effectiveness by the factor 2-4 implies that a climatically significant TSI decrease of roughly 0.15-0.3 % might occur without requiring complete removal of the photospheric magnetic network. *Therefore, significant TSI forcing of climate might have occurred during the prolonged Spörer and Maunder Minima of the $15^{th}$ – $17^{th}$ centuries, while still retaining enough photospheric magnetism to modulate the Be 10 isotope record* (Beer et al. 1998).

f) No significant trend is measured in the network area filling factor over the nine solar minima between 1914 –1996. *So there is no reason to expect 20$^{th}$ century TSI variation beyond that generated by the 11-yr sunspot cycle.*

We thank C. Fröhlich for discussion of the radiometry, and H. Spruit for helpful comments on this paper. The work at Heliophysics, Inc was supported by NASA grants NNX09AP96G and NNX10AC09G.

**References**

Ball,W., Unruh,Y., Krivova,N., Solanki, S., & Harder, J. 2011, A&A, submitted

Beer, J., Tobias, S., & Weiss, N. 1998, Sol. Phys., 181, 237

Berger, T., Schrijver,C., Shine, R., Tarbell, T., Title, A., & Scharmer, G. 1995, ApJ, 454, 531

Carlsson, M., Stein, R., Nordlund, Å., & Scharmer, G. 2004, ApJL, 610, 137

Fligge, M., & Solanki, S. 2000, JA&A, 21, 275

Foukal, P. 1996, Geophys. Res. Lett., 23, 2169

Foukal, P. 1998, Geophys. Res. Lett., 25, 2909

Foukal, P., Bernasconi, P., Eaton, H., & Rust, D. 2004, ApJL, 611, 57

Foukal, P., Bernasconi, P., & Fröhlich, C. 2009, BAAS, 41, #2, 827

Foukal, P., & Fowler, L. 1984, ApJ, 281, 1442

Foukal, P., Fröhlich, C., Spruit, H., & Wigley, T. 2006, Nature, 443, 161

Foukal, P., Harvey, K., & Hill, F. 1991, ApJ, 383, 89

Foukal, P., & Lean, J. 1988, ApJ, 328, 347

Foukal, P., Little, R., Graves, J., Rabin, D., & Lynch, D. 1990, ApJL, 353, 712

Foukal, P., & Milano, L. 2001, Geophys. Res. Lett., 28, 883

Foukal, P., North, J., & Wigley, T. 2004, Science, 306, 68

Fröhlich, C. 2009, A&A, 501, L27

Fröhlich, C., & Lean, J. 2004, A&AR, 12, 273

Hall, J., & Lockwood, W. 2004, ApJ, 614, 942

Keller, C. 1992, Nature, 359, 307

Keller, C., Schüssler, M., Vögler, A., & Zakharov, V. 2004, ApJL, 607, 59

Krivova,N., Solanki,S., & Schmutz,W. 2011, A&A, 529, 81

Lawrence,J.,Chapman,G., & Herzog,A. 1988, ApJ 324,1184

Lawrence, J., Topka, K., & Jones, H. 1993, J. Geophys. Res., 98, 18911

Lean, J., Beer, J., & Bradley, R. 1995, Geophys. Res. Lett., 22, 3195

Lean, J., Skumanich, A., & White, O. 1992, Geophys. Res. Lett., 19, 1591

Lean, J., Wang, Y–M., & Sheeley, N. 2002, Geophys. Res. Lett., 29, 2224

Lites, B., et al. 2004, Sol. Phys., 21, 65

Lockwood, M., & Stamper, R. 1999, Geophys. Res. Lett., 26, 2461

Moran, T., Foukal, P., & Rabin, D. 1992, Sol. Phys., 142, 35

Muller, R. 1975, Sol. Phys., 45, 105

Ortiz, A., Domingo, V., & Sanahuja, B. 2006, A&A, 452, 311

Schnerr, R., & Spruit, H. 2010, A&A, submitted

Schussler, M., & Vogler, A., 2008, A&A 481,L5

Skumanich, A., Smythe, C., & Frazier, E. 1975, ApJ, 200, 747

Solanki, S., & Fligge, M. 1998, Geophys. Res. Lett., 25, 341

Solanki, S., & Krivova, N. 2004, Sol. Phys., 224, 197

Spruit, H. 1976, Sol, Phys., 50, 269

Spruit, H., & Zwaan, C. 1981, Sol. Phys., 70, 207

Steinhilber, F. 2010, A&A, 523, A39

Stenflo, J. 1973, Sol. Phys., 32, 41

Svalgaard, L., & Cliver, E. 2010, J. Geophys. Res., 115, AO9111, doi: 10.1029/2009JA015069 2010

Wang, Y., Sheeley, N., & Lean, J. 2000, Geophys. Res. Lett., 27, 621

Wang, H., Spirock, T., Goode, P., Lee, C., Zirin, H., & Kosonocky, W. 1998, ApJ, 495, 957

Wigley, T., & Raper, S. 1990, Geophys. Res. Lett., 17, 2169

Worden, P. 1975, Sol. Phys., 45, 52

Zirin, H., & Wang, H. 1992, ApJL, 385, 27

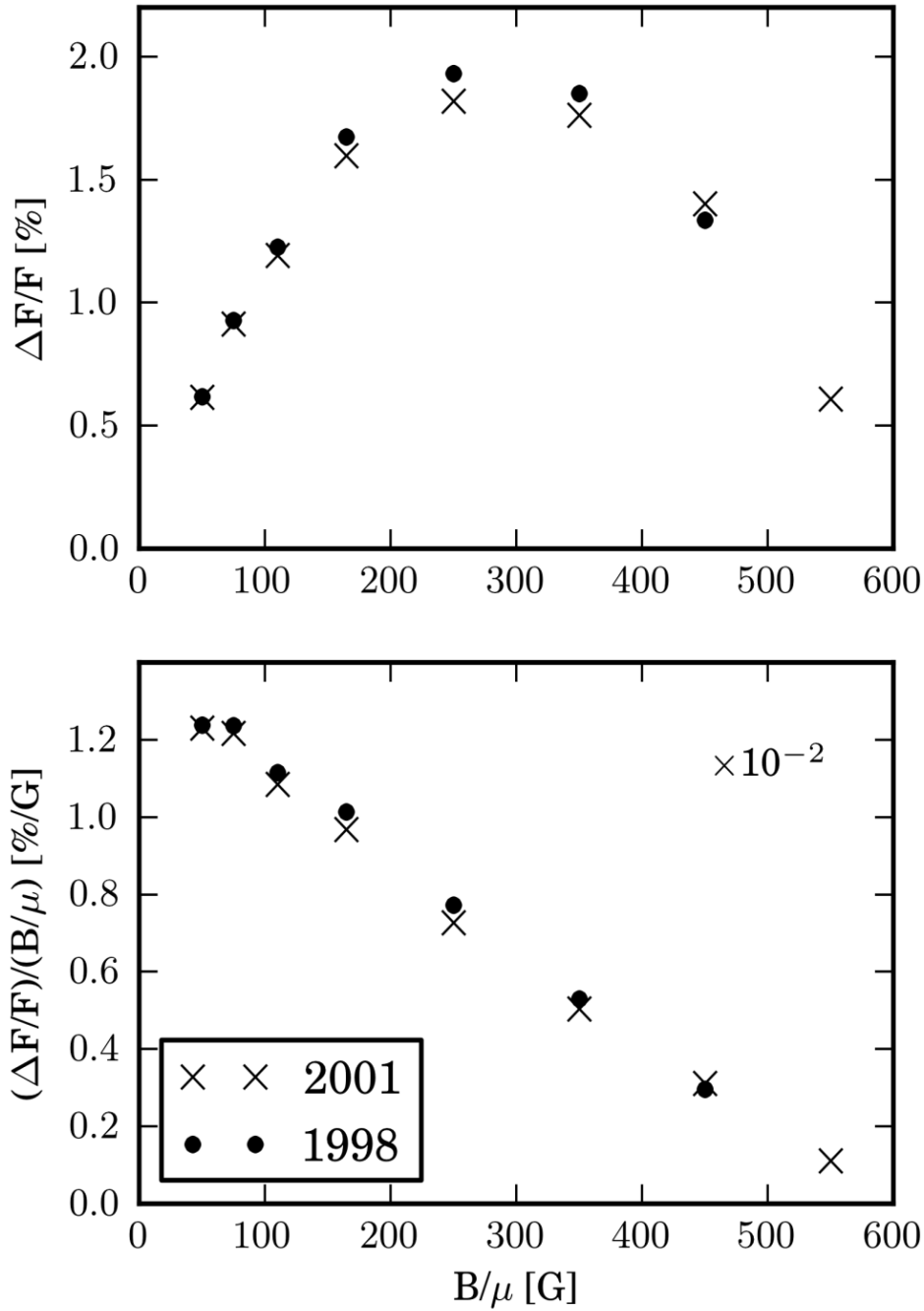

Figure 1.Upper panel: Normalized excess radiative fluxes, ΔF/F, of faculae (ordinate) plotted against their magnetic signal, B/μ, corrected for projection. Lower panel: Normalized excess radiative fluxes / corrected magnetograph signal versus corrected magnetograph signal. Crosses and dots denote data obtained at high (2001) and lower (1998) solar activity.

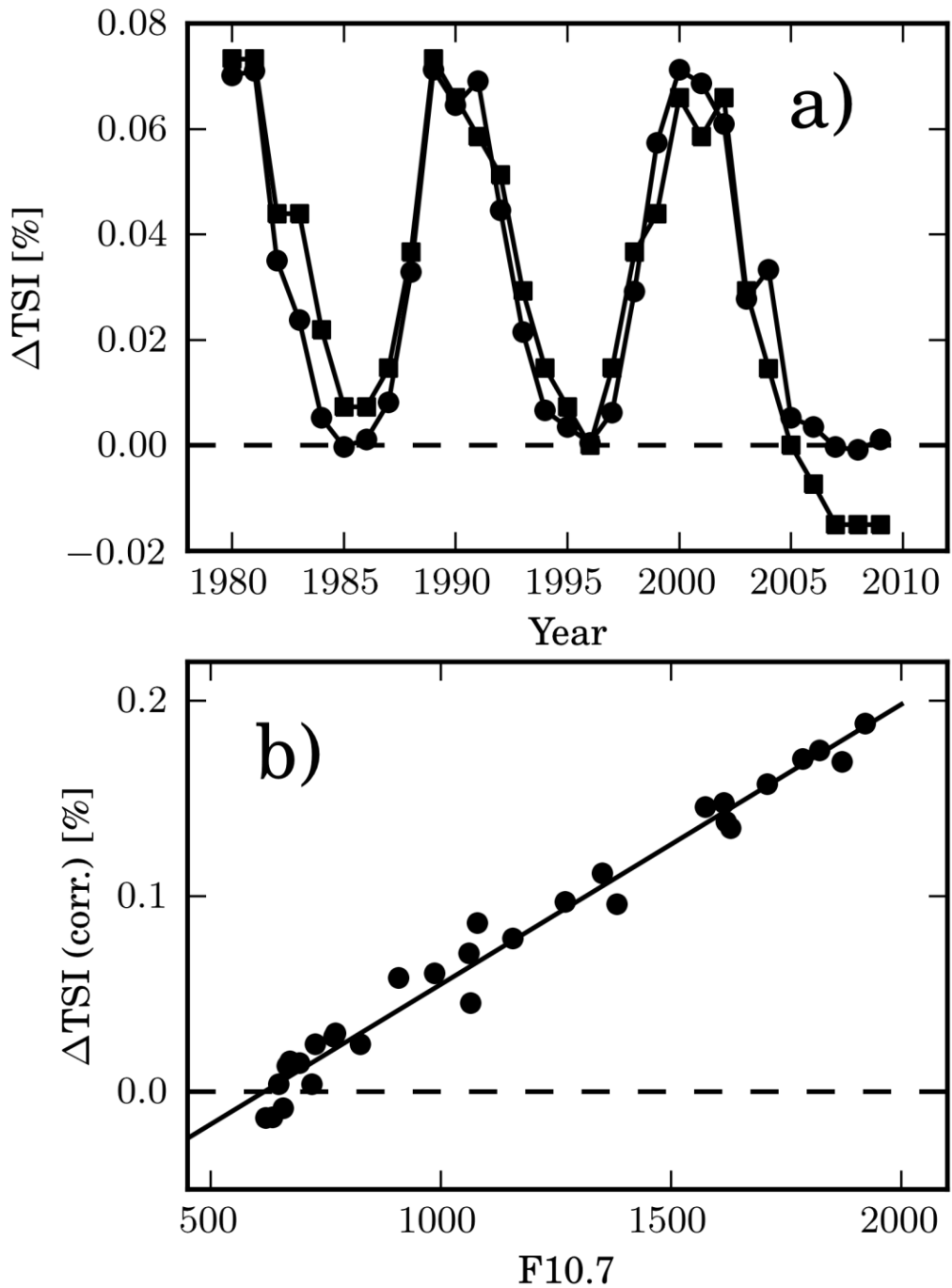

Figure 2. Panel (a): Annual mean PMOD radiometry (squares), and TSI reconstructed (filled circles). Panel (b): Linear regression of PMOD TSI (corrected for sunspot blocking) versus F10.7.